\documentstyle[12pt,epsf]{article}

\makeatletter
\newcount\GRAPHICSTYPE
\GRAPHICSTYPE=\z@
\def\GRAPHICSPS#1{%
 \ifcase\GRAPHICSTYPE
  ps: #1%
 \or
  language "PS", include "#1"%
 \fi
}%
\def\graffile#1#2#3#4{%
    \leavevmode
    \raise -#4 \BOXTHEFRAME{%
        \hbox to #2{\raise #3\hbox{\null #1}}}%
}%
\def\draftbox#1#2#3#4{%
 \leavevmode\raise -#4 \hbox{%
  \frame{\rlap{\protect\tiny #1}\hbox to #2%
   {\vrule height#3 width\z@ depth\z@\hfil}%
  }%
 }%
}%
\newcount\draft
\draft=\z@
\def\GRAPHIC#1#2#3#4#5{%
 \ifnum\draft=\@ne\draftbox{#2}{#3}{#4}{#5}%
  \else\graffile{#1}{#3}{#4}{#5}%
  \fi
 }%
\def\addtoLaTeXparams#1{%
    \edef\LaTeXparams{\LaTeXparams #1}}%
\newif\ifBoxFrame \BoxFramefalse
\newif\ifOverFrame \OverFramefalse

\def\BOXTHEFRAME#1{%
   \hbox{%
      \ifBoxFrame
         \frame{#1}%
      \else
         {#1}%
      \fi
   }%
}

\def\doFRAMEparams#1{\BoxFramefalse\OverFramefalse\readFRAMEparams#1\end}%
\def\readFRAMEparams#1{%
 \ifx#1\end%
  \let\next=\relax
  \else
  \ifx#1i\dispkind=\z@\fi
  \ifx#1d\dispkind=\@ne\fi
  \ifx#1f\dispkind=\tw@\fi
  \ifx#1t\addtoLaTeXparams{t}\fi
  \ifx#1b\addtoLaTeXparams{b}\fi
  \ifx#1p\addtoLaTeXparams{p}\fi
  \ifx#1h\addtoLaTeXparams{h}\fi
  \ifx#1X\BoxFrametrue\fi
  \ifx#1O\OverFrametrue\fi
  \let\next=\readFRAMEparams
  \fi
 \next
 }%
\def\IFRAME#1#2#3#4#5#6{%
      \bgroup
      \parindent=0pt%
      \setbox0 = \hbox{#6}%
      \@tempdima = #1%
      \ifOverFrame
          \typeout{This is not implemented yet}%
          \show\HELP
      \else
         \ifdim\wd0>\@tempdima
            \advance\@tempdima by \@tempdima
            \ifdim\wd0 >\@tempdima
               \textwidth=\@tempdima
               \setbox1 =\vbox{%
                  \noindent\hbox to
\@tempdima{\hfill\GRAPHIC{#5}{#4}{#1}{#2}{#3}\hfill}\\%
                  \noindent\hbox to \@tempdima{\parbox[b]{\@tempdima}{#6}}%
               }%
               \wd1=\@tempdima
            \else
               \textwidth=\wd0
               \setbox1 =\vbox{%
                 \noindent\hbox to
\wd0{\hfill\GRAPHIC{#5}{#4}{#1}{#2}{#3}\hfill}\\%
                 \noindent\hbox{#6}%
               }%
               \wd1=\wd0
            \fi
         \else
            \hsize=\@tempdima
            \setbox1 =\vbox{%
                \unskip\GRAPHIC{#5}{#4}{#1}{#2}{0pt}%
                \break
                \unskip\hbox to \@tempdima{\hfill #6\hfill}%
            }%
            \wd1=\@tempdima
         \fi
         \@tempdimb=\ht1
         \advance\@tempdimb by \dp1
         \advance\@tempdimb by -#2%
         \advance\@tempdimb by #3%
         \leavevmode
         \raise -\@tempdimb \hbox{\box1}%
      \fi
      \egroup
}%
\def\DFRAME#1#2#3#4#5{%
 \begin{center}
     \ifOverFrame
        #5\par
     \fi
     \GRAPHIC{#4}{#3}{#1}{#2}{\z@}
     \ifOverFrame \else
        \par #5
     \fi
 \end{center}%
 }%
\def\FFRAME#1#2#3#4#5#6#7{%
 \begin{figure}[#1]%
  \begin{center}\GRAPHIC{#7}{#6}{#2}{#3}{\z@}\end{center}%
  \caption{\label{#5}#4}%
  \end{figure}%
 }%
\newcount\dispkind%
\def\FRAME#1#2#3#4#5#6#7#8{%
 \def\LaTeXparams{}%
 \dispkind=\z@
 \def\LaTeXparams{}%
 \doFRAMEparams{#1}%
 \ifnum\dispkind=\z@\IFRAME{#2}{#3}{#4}{#7}{#8}{#5}\else
  \ifnum\dispkind=\@ne\DFRAME{#2}{#3}{#7}{#8}{#5}\else
   \ifnum\dispkind=\tw@
    \edef\@tempa{\noexpand\FFRAME{\LaTeXparams}}%
    \@tempa{#2}{#3}{#5}{#6}{#7}{#8}%
    \fi
   \fi
  \fi
 }%
\def\TEXUX#1{"texux"}
\def\@@eqncr{\let\@tempa\relax
    \ifcase\@eqcnt \def\@tempa{& & &}\or \def\@tempa{& &}%
      \else \def\@tempa{&}\fi
     \@tempa
     \if@eqnsw
        \iftag@
           \@taggnum
        \else
           \@eqnnum\stepcounter{equation}\fi
     \fi
     \global\tag@false
     \global\@eqnswtrue
     \global\@eqcnt\z@\cr}

 \newif\iftag@ \tag@false
 \def\tag{\@ifnextchar*{\@tagstar}{\@tag}}
 \def\@tag#1{%
     \global\tag@true
     \global\def\@taggnum{(#1)}}
 \def\@tagstar*#1{%
     \global\tag@true
     \global\def\@taggnum{#1}%
}
\long\def\QQQ#1#2{%
     \long\expandafter\def\csname#1\endcsname{#2}}%
\@ifundefined{QTP}{\def\QTP#1{}}{}
\@ifundefined{Qcb}{}{}
\@ifundefined{Qct}{}{}
\@ifundefined{Qlb}{}{}
\@ifundefined{Qlt}{}{}
\long\def\QQA#1#2{}%
\def\QTR#1#2{{\csname#1\endcsname #2}}
\def\EXPAND#1[#2]#3{}%
\def\NOEXPAND#1[#2]#3{}%
\def\LaTeXparent#1{}%
\def\ChildStyles#1{}%
\def\ChildDefaults#1{}%
\def\QTagDef#1#2#3{}%
\def\QQfnmark#1{\footnotemark}

\@ifundefined{INDEX}{\def\INDEX#1#2{}{}}{}%
\@ifundefined{SUBINDEX}{\def\SUBINDEX#1#2#3{}{}{}}{}%
\def\initial#1{\bigbreak{\raggedright\large\bf #1}\kern 2\p@
   \penalty3000}%
\def\entry#1#2{\item {#1}, #2}%
\@ifundefined{ZZZ}{}{\makeatletter\input gnuindex.sty\makeatother\makeindex\makeatletter}%
\@ifundefined{abstract}{%
 \def\abstract{%
  \if@twocolumn
   \section*{Abstract (Not appropriate in this style!)}%
   \else \small
   \begin{center}{\bf Abstract\vspace{-.5em}\vspace{\z@}}\end{center}%
   \quotation
   \fi
  }%
 }{%
 }%
\@ifundefined{endabstract}{\def\endabstract
  {\if@twocolumn\else\endquotation\fi}}{}%
\@ifundefined{maketitle}{\def\maketitle#1{}}{}%
\@ifundefined{affiliation}{\def\affiliation#1{}}{}%
\@ifundefined{proof}{}{}%
\@ifundefined{endproof}{}{}%
\@ifundefined{newfield}{\def\newfield#1#2{}}{}%
\@ifundefined{chapter}{\def\chapter#1{\par(Chapter head:)#1\par }%
 \newcount\c@chapter}{}%
\@ifundefined{part}{\def\part#1{\par(Part head:)#1\par }}{}%
\@ifundefined{section}{\def\section#1{\par(Section head:)#1\par }}{}%
\@ifundefined{subsection}{\def\subsection#1%
 {\par(Subsection head:)#1\par }}{}%
\@ifundefined{subsubsection}{\def\subsubsection#1%
 {\par(Subsubsection head:)#1\par }}{}%
\@ifundefined{paragraph}{\def\paragraph#1%
 {\par(Subsubsubsection head:)#1\par }}{}%
\@ifundefined{subparagraph}{\def\subparagraph#1%
 {\par(Subsubsubsubsection head:)#1\par }}{}%
\@ifundefined{therefore}{}{}%
\@ifundefined{backepsilon}{}{}%
\@ifundefined{yen}{}{}%
\@ifundefined{registered}{%
   \def\registered{\relax\ifmmode{}\r@gistered
                    \else$\m@th\r@gistered$\fi}%
 \def\r@gistered{^{\ooalign
  {\hfil\raise.07ex\hbox{$\scriptstyle\rm\text{R}$}\hfil\crcr
  \mathhexbox20D}}}}{}%
\@ifundefined{Eth}{}{}%
\@ifundefined{eth}{}{}%
\@ifundefined{Thorn}{}{}%
\@ifundefined{thorn}{}{}%
\@ifundefined{degree}{}{}%
\def\BibTeX{{\rm B\kern-.05em{\sc i\kern-.025em b}\kern-.08em
    T\kern-.1667em\lower.7ex\hbox{E}\kern-.125emX}}%
\newdimen\theight
\def\Column{%
 \vadjust{\setbox\z@=\hbox{\scriptsize\quad\quad tcol}%
  \theight=\ht\z@\advance\theight by \dp\z@\advance\theight by \lineskip
  \kern -\theight \vbox to \theight{%
   \rightline{\rlap{\box\z@}}%
   \vss
   }%
  }%
 }%
\def\qed{%
 \ifhmode\unskip\nobreak\fi\ifmmode\ifinner\else\hskip5\p@\fi\fi
 \hbox{\hskip5\p@\vrule width4\p@ height6\p@ depth1.5\p@\hskip\p@}%
 }%
\def\miss{\hbox{\vrule height2\p@ width 2\p@ depth\z@}}%
\def\tcol#1{{\baselineskip=6\p@ \vcenter{#1}} \Column}  %
\def\newfmtname{LaTeX2e}
\def\chkcompat{%
   \if@compatibility
   \else
     \usepackage{latexsym}
   \fi
}

\ifx\fmtname\newfmtname
  \DeclareOldFontCommand{\rm}{\normalfont\rmfamily}{\mathrm}
  \DeclareOldFontCommand{\sf}{\normalfont\sffamily}{\mathsf}
  \DeclareOldFontCommand{\tt}{\normalfont\ttfamily}{\mathtt}
  \DeclareOldFontCommand{\bf}{\normalfont\bfseries}{\mathbf}
  \DeclareOldFontCommand{\it}{\normalfont\itshape}{\mathit}
  \DeclareOldFontCommand{\sl}{\normalfont\slshape}{\@nomath\sl}
  \DeclareOldFontCommand{\sc}{\normalfont\scshape}{\@nomath\sc}
  \chkcompat
\fi

\def\alpha{\Greekmath 010B }%
\def\beta{\Greekmath 010C }%
\def\gamma{\Greekmath 010D }%
\def\delta{\Greekmath 010E }%
\def\epsilon{\Greekmath 010F }%
\def\zeta{\Greekmath 0110 }%
\def\eta{\Greekmath 0111 }%
\def\theta{\Greekmath 0112 }%
\def\iota{\Greekmath 0113 }%
\def\kappa{\Greekmath 0114 }%
\def\lambda{\Greekmath 0115 }%
\def\mu{\Greekmath 0116 }%
\def\nu{\Greekmath 0117 }%
\def\xi{\Greekmath 0118 }%
\def\pi{\Greekmath 0119 }%
\def\rho{\Greekmath 011A }%
\def\sigma{\Greekmath 011B }%
\def\tau{\Greekmath 011C }%
\def\upsilon{\Greekmath 011D }%
\def\phi{\Greekmath 011E }%
\def\chi{\Greekmath 011F }%
\def\psi{\Greekmath 0120 }%
\def\omega{\Greekmath 0121 }%
\def\varepsilon{\Greekmath 0122 }%
\def\vartheta{\Greekmath 0123 }%
\def\varpi{\Greekmath 0124 }%
\def\varrho{\Greekmath 0125 }%
\def\varsigma{\Greekmath 0126 }%
\def\varphi{\Greekmath 0127 }%

\def\nabla{\Greekmath 0272}

\def\GreekBold{\@ne}%
\def\One{\@ne}

\def\Greekmath#1#2#3#4{%
    \ifx\GreekBold\One
        \mathchar"#1#2#3#4%
    \else
		\mbox{\boldmath$\mathchar"#1#2#3#4$}
	\fi}

\let\SAVEPBF=\pbf

\def\pbf{\let\GreekBold = \relax\SAVEPBF}%

\expandafter\ifx\csname ds@amstex\endcsname\relax
\else\makeatother\endinput\fi

\let\DOTSI\relax
\def\RIfM@{\relax\ifmmode}%
\def\FN@{\futurelet\next}%
\newcount\intno@
\def\iint{\DOTSI\intno@\tw@\FN@\ints@}%
\def\iiint{\DOTSI\intno@\thr@@\FN@\ints@}%
\def\iiiint{\DOTSI\intno@4 \FN@\ints@}%
\def\idotsint{\DOTSI\intno@\z@\FN@\ints@}%
\def\ints@{\findlimits@\ints@@}%
\newif\iflimtoken@
\newif\iflimits@
\def\findlimits@{\limtoken@true\ifx\next\limits\limits@true
 \else\ifx\next\nolimits\limits@false\else
 \limtoken@false\ifx\ilimits@\nolimits\limits@false\else
 \ifinner\limits@false\else\limits@true\fi\fi\fi\fi}%
\def\multint@{\int\ifnum\intno@=\z@\intdots@                          
 \else\intkern@\fi                                                    
 \ifnum\intno@>\tw@\int\intkern@\fi                                   
 \ifnum\intno@>\thr@@\int\intkern@\fi                                 
 \int}
\def\multintlimits@{\intop\ifnum\intno@=\z@\intdots@\else\intkern@\fi
 \ifnum\intno@>\tw@\intop\intkern@\fi
 \ifnum\intno@>\thr@@\intop\intkern@\fi\intop}%
\def\intic@{%
    \mathchoice{\hskip.5em}{\hskip.4em}{\hskip.4em}{\hskip.4em}}%
\def\negintic@{\mathchoice
 {\hskip-.5em}{\hskip-.4em}{\hskip-.4em}{\hskip-.4em}}%
\def\ints@@{\iflimtoken@                                              
 \def\ints@@@{\iflimits@\negintic@
   \mathop{\intic@\multintlimits@}\limits                             
  \else\multint@\nolimits\fi                                          
  \eat@}
 \else                                                                
 \def\ints@@@{\iflimits@\negintic@
  \mathop{\intic@\multintlimits@}\limits\else
  \multint@\nolimits\fi}\fi\ints@@@}%
\def\intkern@{\mathchoice{\!\!\!}{\!\!}{\!\!}{\!\!}}%
\def\plaincdots@{\mathinner{\cdotp\cdotp\cdotp}}%
\def\intdots@{\mathchoice{\plaincdots@}%
 {{\cdotp}\mkern1.5mu{\cdotp}\mkern1.5mu{\cdotp}}%
 {{\cdotp}\mkern1mu{\cdotp}\mkern1mu{\cdotp}}%
 {{\cdotp}\mkern1mu{\cdotp}\mkern1mu{\cdotp}}}%
\def\RIfM@{\relax\protect\ifmmode}
\def\text{\RIfM@\expandafter\text@\else\expandafter\mbox\fi}
\let\nfss@text\text
\def\text@#1{\mathchoice
   {\textdef@\displaystyle\f@size{#1}}%
   {\textdef@\textstyle\tf@size{\firstchoice@false #1}}%
   {\textdef@\textstyle\sf@size{\firstchoice@false #1}}%
   {\textdef@\textstyle \ssf@size{\firstchoice@false #1}}%
   \glb@settings}

\def\textdef@#1#2#3{\hbox{{%
                    \everymath{#1}%
                    \let\f@size#2\selectfont
                    #3}}}
\newif\iffirstchoice@
\firstchoice@true
\def\Let@{\relax\iffalse{\fi\let\\=\cr\iffalse}\fi}%
\def\vspace@{\def\vspace##1{\crcr\noalign{\vskip##1\relax}}}%
\def\multilimits@{\bgroup\vspace@\Let@
 \baselineskip\fontdimen10 \scriptfont\tw@
 \advance\baselineskip\fontdimen12 \scriptfont\tw@
 \lineskip\thr@@\fontdimen8 \scriptfont\thr@@
 \lineskiplimit\lineskip
 \vbox\bgroup\ialign\bgroup\hfil$\m@th\scriptstyle{##}$\hfil\crcr}%
\def\Sb{_\multilimits@}%
\def\endSb{\crcr\egroup\egroup\egroup}%
\def\Sp{^\multilimits@}%

\newdimen\ex@
\def\rightarrowfill@#1{$#1\m@th\mathord-\mkern-6mu\cleaders
 \hbox{$#1\mkern-2mu\mathord-\mkern-2mu$}\hfill
 \mkern-6mu\mathord\rightarrow$}%
\def\leftarrowfill@#1{$#1\m@th\mathord\leftarrow\mkern-6mu\cleaders
 \hbox{$#1\mkern-2mu\mathord-\mkern-2mu$}\hfill\mkern-6mu\mathord-$}%
\def\leftrightarrowfill@#1{$#1\m@th\mathord\leftarrow
\mkern-6mu\cleaders
 \hbox{$#1\mkern-2mu\mathord-\mkern-2mu$}\hfill
 \mkern-6mu\mathord\rightarrow$}%
\def\overrightarrow{\mathpalette\overrightarrow@}%
\def\overrightarrow@#1#2{\vbox{\ialign{##\crcr\rightarrowfill@#1\crcr
 \noalign{\kern-\ex@\nointerlineskip}$\m@th\hfil#1#2\hfil$\crcr}}}%

\def\overleftarrow{\mathpalette\overleftarrow@}%
\def\overleftarrow@#1#2{\vbox{\ialign{##\crcr\leftarrowfill@#1\crcr
 \noalign{\kern-\ex@\nointerlineskip}$\m@th\hfil#1#2\hfil$\crcr}}}%
\def\overleftrightarrow{\mathpalette\overleftrightarrow@}%
\def\overleftrightarrow@#1#2{\vbox{\ialign{##\crcr
   \leftrightarrowfill@#1\crcr
 \noalign{\kern-\ex@\nointerlineskip}$\m@th\hfil#1#2\hfil$\crcr}}}%
\def\underrightarrow{\mathpalette\underrightarrow@}%
\def\underrightarrow@#1#2{\vtop{\ialign{##\crcr$\m@th\hfil#1#2\hfil
  $\crcr\noalign{\nointerlineskip}\rightarrowfill@#1\crcr}}}%

\def\underleftarrow{\mathpalette\underleftarrow@}%
\def\underleftarrow@#1#2{\vtop{\ialign{##\crcr$\m@th\hfil#1#2\hfil
  $\crcr\noalign{\nointerlineskip}\leftarrowfill@#1\crcr}}}%
\def\underleftrightarrow{\mathpalette\underleftrightarrow@}%
\def\underleftrightarrow@#1#2{\vtop{\ialign{##\crcr$\m@th
  \hfil#1#2\hfil$\crcr
 \noalign{\nointerlineskip}\leftrightarrowfill@#1\crcr}}}%

\def\qopnamewl@#1{\mathop{\operator@font#1}\nlimits@}
\let\nlimits@\displaylimits
\def\setboxz@h{\setbox\z@\hbox}

\def\varlim@#1#2{\mathop{\vtop{\ialign{##\crcr
 \hfil$#1\m@th\operator@font lim$\hfil\crcr
 \noalign{\nointerlineskip}#2#1\crcr
 \noalign{\nointerlineskip\kern-\ex@}\crcr}}}}

 \def\rightarrowfill@#1{\m@th\setboxz@h{$#1-$}\ht\z@\z@
  $#1\copy\z@\mkern-6mu\cleaders
  \hbox{$#1\mkern-2mu\box\z@\mkern-2mu$}\hfill
  \mkern-6mu\mathord\rightarrow$}
\def\leftarrowfill@#1{\m@th\setboxz@h{$#1-$}\ht\z@\z@
  $#1\mathord\leftarrow\mkern-6mu\cleaders
  \hbox{$#1\mkern-2mu\copy\z@\mkern-2mu$}\hfill
  \mkern-6mu\box\z@$}

\def\projlim{\qopnamewl@{proj\,lim}}
\def\injlim{\qopnamewl@{inj\,lim}}
\def\varinjlim{\mathpalette\varlim@\rightarrowfill@}
\def\varprojlim{\mathpalette\varlim@\leftarrowfill@}
\def\varliminf{\mathpalette\varliminf@{}}
\def\varliminf@#1{\mathop{\underline{\vrule\@depth.2\ex@\@width\z@
   \hbox{$#1\m@th\operator@font lim$}}}}
\def\varlimsup{\mathpalette\varlimsup@{}}
\def\varlimsup@#1{\mathop{\overline
  {\hbox{$#1\m@th\operator@font lim$}}}}

\begingroup \catcode `|=0 \catcode `[= 1
\catcode`]=2 \catcode `\{=12 \catcode `\}=12
\catcode`\\=12
|gdef|@alignverbatim#1\end{align}[#1|end[align]]
|gdef|@salignverbatim#1\end{align*}[#1|end[align*]]

|gdef|@alignatverbatim#1\end{alignat}[#1|end[alignat]]
|gdef|@salignatverbatim#1\end{alignat*}[#1|end[alignat*]]

|gdef|@xalignatverbatim#1\end{xalignat}[#1|end[xalignat]]
|gdef|@sxalignatverbatim#1\end{xalignat*}[#1|end[xalignat*]]

|gdef|@gatherverbatim#1\end{gather}[#1|end[gather]]
|gdef|@sgatherverbatim#1\end{gather*}[#1|end[gather*]]

|gdef|@gatherverbatim#1\end{gather}[#1|end[gather]]
|gdef|@sgatherverbatim#1\end{gather*}[#1|end[gather*]]

|gdef|@multilineverbatim#1\end{multiline}[#1|end[multiline]]
|gdef|@smultilineverbatim#1\end{multiline*}[#1|end[multiline*]]

|gdef|@arraxverbatim#1\end{arrax}[#1|end[arrax]]
|gdef|@sarraxverbatim#1\end{arrax*}[#1|end[arrax*]]

|gdef|@tabulaxverbatim#1\end{tabulax}[#1|end[tabulax]]
|gdef|@stabulaxverbatim#1\end{tabulax*}[#1|end[tabulax*]]

|endgroup

\def\align{\@verbatim \frenchspacing\@vobeyspaces \@alignverbatim
You are using the "align" environment in a style in which it is not defined.}

\@namedef{align*}{\@verbatim\@salignverbatim
You are using the "align*" environment in a style in which it is not defined.}
\expandafter\let\csname endalign*\endcsname =\endtrivlist

\def\alignat{\@verbatim \frenchspacing\@vobeyspaces \@alignatverbatim
You are using the "alignat" environment in a style in which it is not defined.}

\@namedef{alignat*}{\@verbatim\@salignatverbatim
You are using the "alignat*" environment in a style in which it is not
defined.}
\expandafter\let\csname endalignat*\endcsname =\endtrivlist

\def\xalignat{\@verbatim \frenchspacing\@vobeyspaces \@xalignatverbatim
You are using the "xalignat" environment in a style in which it is not
defined.}

\@namedef{xalignat*}{\@verbatim\@sxalignatverbatim
You are using the "xalignat*" environment in a style in which it is not
defined.}
\expandafter\let\csname endxalignat*\endcsname =\endtrivlist

\def\gather{\@verbatim \frenchspacing\@vobeyspaces \@gatherverbatim
You are using the "gather" environment in a style in which it is not defined.}

\@namedef{gather*}{\@verbatim\@sgatherverbatim
You are using the "gather*" environment in a style in which it is not defined.}
\expandafter\let\csname endgather*\endcsname =\endtrivlist

\def\multiline{\@verbatim \frenchspacing\@vobeyspaces \@multilineverbatim
You are using the "multiline" environment in a style in which it is not
defined.}

\@namedef{multiline*}{\@verbatim\@smultilineverbatim
You are using the "multiline*" environment in a style in which it is not
defined.}
\expandafter\let\csname endmultiline*\endcsname =\endtrivlist

\def\arrax{\@verbatim \frenchspacing\@vobeyspaces \@arraxverbatim
You are using a type of "array" construct that is only allowed in AmS-LaTeX.}

\def\tabulax{\@verbatim \frenchspacing\@vobeyspaces \@tabulaxverbatim
You are using a type of "tabular" construct that is only allowed in AmS-LaTeX.}

\@namedef{arrax*}{\@verbatim\@sarraxverbatim
You are using a type of "array*" construct that is only allowed in AmS-LaTeX.}
\expandafter\let\csname endarrax*\endcsname =\endtrivlist

\@namedef{tabulax*}{\@verbatim\@stabulaxverbatim
You are using a type of "tabular*" construct that is only allowed in
AmS-LaTeX.}
\expandafter\let\csname endtabulax*\endcsname =\endtrivlist

\@ifundefined{theorem}{}{}
\@ifundefined{lemma}{}{}
\@ifundefined{corollary}{}{}
\@ifundefined{conjecture}{}{}
\@ifundefined{proposition}{}{}
\@ifundefined{axiom}{}{}
\@ifundefined{remark}{}{}
\@ifundefined{example}{}{}
\@ifundefined{exercise}{}{}
\@ifundefined{definition}{}{}

\makeatother

    \def\initial#1{\bigbreak{\raggedright\large\bf #1}\kern 2pt\penalty3000}
    \def\entry#1#2{\item {#1}, #2}

  \def\INDEX{\@bsphack\begingroup\@sanitize\@WRINDEX\@indexfile}
  \def\@WRINDEX#1#2#3{\let\thepage\relax
     \xdef\@gtempa{\write#1{\string
      \entry{#2}{\thepage}{#3}}}\endgroup\@gtempa
     \if@nobreak \ifvmode\nobreak\fi\fi\@esphack}
  \def\SUBINDEX{\@bsphack\begingroup\@sanitize\@WRSUBINDEX\@indexfile}
  \def\@WRSUBINDEX#1#2#3#4{\let\thepage\relax
     \xdef\@gtempa{\write#1{\string
     \entry{#2\space\space\space#4}{\thepage}{#3}{#4}}}\endgroup\@gtempa
     \if@nobreak \ifvmode\nobreak\fi\fi\@esphack}

\QQQ{Language}{
American English
}

\begin{document}


\title{
{\small{hep-th/9412044 \hfill USC-94/HEP-B4} } \\
Folded Strings\thanks{%
Based on a lecture delivered at the G\"ursey Memorial Conference I,
Strings and Symmetries, Istanbul, June 1994. To appear in the proceedings.}}
\author{Itzhak Bars\thanks{%
Research supported in part by the Deparment of Energy under Grant No.
DE-FG03-84ER40168.} \\
Department of Physics and Astronomy\\
University of Southern California\\
Los Angeles, CA 90089-0484}
\date{}
\maketitle

\begin{abstract}
Recent progress on the complete set of solutions of two dimensional
classical string theory in any curved spacetime is reviewed. When the
curvature is smooth, the string solutions are deformed folded string
solutions as compared to flat spacetime folded strings that were known for
19 years. However, surprizing new stringy behavior becomes evident at
singularities such as black holes. The global properties of these solutions
require that the ``bare singularity region'' of the black hole be included
along with the usual black hole spacetime. The mathematical structure needed
to describe the solutions include a recursion relation that is analogous to
the transfer matrix of lattice theories. This encodes lattice properties on
the worldsheet on the one hand and the geometry of spacetime on the other
hand. A case is made for the presence of folded strings in the quantum
theory of non-critical strings for $d\geq 2$.


\end{abstract}

\newpage
\section{Introduction}

Feza G\"ursey was a great master and an artist in finding connections
between Mathematics and Physics. In this conference we have the pleasure to
hear from many of his friends that have admired his leadership in several
areas of Physics. I am very appreciative for having been given the
opportunity to express my gratitude to Feza for the inspiration he has
provided to me as my teacher as well as my colleague and friend.

Among his first discoveries was the {\it non-linear sigma models}, which he
applied to pion physics\footnote{%
Feza was the first to introduce the idea of the sigma models. Many people
think of the paper by Gell-Mann and L\'evy in connection with sigma models,
but it is important to recall that Gell-Mann and L\'evy refer to Feza's
paper.}. Nowdays, sigma models are at the basis of string theory in the form
of conformal field theories. In recent years, through the use of gauged
Wess-Zumino-Witten models based on non-compact groups, it has been possible
to construct {\it exact conformal field theories} that describe
(super)strings propagating in curved space-time in 2D to 4D. These models
combine several fields that deeply interested Feza: non-linear sigma models,
conformal invariance, classical and quantum gravity, non-compact groups,
unification of forces, string theory. In honoring Feza today, I would like
to highlight recent progress made in this field.

\section{Motivation}

The original physical motivations for studying string theory were: (1)
understanding unification of forces including quantum gravity, and (2)
understanding the Standard Model. In recent years it has become more and
more evident that these goals should be examined in the presence of curved
4D space-time string backgrounds. The construction of 4D curved space-time
string theories that correspond to exact conformal theories have provided
models in which various questions can be investigated \cite{ibcurved}\cite
{tseytlinrev}.

The usual scenario of flat 4D plus extra curved dimensions may not be the
right approach for making predictions about the Standard Model. The gauge
symmetries and spectrum of quark + lepton families, which are the main
ingredients of the Standard Model, were probably fixed during the early
times in the evolution of the Universe. At such times 4D space-time was
curved. Since curvature contributes to the central charge and other
topological aspects of String Theory, it is likely that the predictions of
String Theory under such conditions may be quite different than the flat 4D
approach. Therefore, String Theory in curved space-time must be better
understood before attempting to make connections to low energy physics. One
should consider all kinds of curved backgrounds, not only the traditional
cosmological backgrounds, since the passage from curved space-time to flat
space-time may involve various phase transitions, including inflation of a
small region of the original curved universe to today's universe that is
essentially homogeneous and flat. The gauge bosons, and chiral families of
quarks and leptons in a small region of the early curved universe would
become the ones observed in today's inflated flat universe. The possibility
of such a scenario suggests that curved space-time string theory deserves
intensive study. In addition, the issues surrounding gravitational
singularities should be answered in the context of curved space-time string
theory, as it is the only known theory of quantum gravity.

\section{Some Results in 1-time G/H}

With these questions in mind, we have been pursuing a program of building
and analyzing exactly solvable models of string theory in curved spacetime
based on conformal field theory. The main tool is the G/H gauged WZW model
based on non-compact groups, such that the coset contains a single time
coordinate. A lot of progress was made on the construction of exact
conformal field theory models for bosonic, supersymmetric and heterotic
strings in curved spacetime, and some exact results were derived. These
include :

\begin{itemize}
\item  Classification of G/H models with 1-time + ($d-1)$-space coordinates
\cite{ibcurved}. G is non-compact and H can be non-compact or compact. After
identifying the simple cases for G/H that yield a single time coordinate,
the classification is easily extended to semi-simple, with Abelian factors,
and their contractions to solvable groups \cite{witnapi}\cite{sfrecent}.
There may also exist other exact conformal models which may not be G/H
models.

\item  The non-linear sigma model geometries for these models have been
derived \cite{ibsfglo}\cite{ibsfexa}\cite{ibcurved}\cite{tseytlinrev},
giving the metric $G_{\mu \nu }(x)$, the anti-symmetric tensor $B_{\mu \nu
}(x)$, and the dilaton $\Phi (x).$ These automatically solve Einstein's
equations for dilaton gravity. The global spaces for these manifolds have
been constructed, and rich duality symmetries have been identified.
Furthermore, the exact point particle geodesics in the global space have
been obtained through group theoretical methods \cite{ibsfglo}\cite{ibsfexa}%
\cite{ibcurved}.

\item  Quantum corrections to these geometries have been computed to all
orders in the sigma model interactions, thanks to the group theoretical
construction, mainly by using algebraic methods\cite{ibsfexa}. This led to
an exact quantum effective action \cite{ibsfeffaction} with the quantum
corrected metric, antisymmetric tensor and dilaton. The results show that
for type-II superstrings, thanks to the supersymmetry, there are no
corrections to the classical expressions. In bosonic or heterotic cases the
corrections show that certain singularities of the metric get shielded by
quantum corrections in parts of spacetime.

\item  In these models the exact spectrum of the Laplacian can be obtained
through unitary representation theory of non-compact groups. This provides
the method for extracting the spectrum of quarks and leptons, but more work
is needed along these lines.
\end{itemize}

\section{Classical solutions}

More recently, it became apparent that a physical interpretation of the
models as well as further progress will be accomplished through the study of
the classical equations of such models. Therefore, we have turned to the
classical theory. This is relevant to fundamental questions of singularities
in gravitational physics, as well as stringy questions about the early
universe and its influence on the low energy spectrum of quarks and leptons.
The classical string solution for any gauged WZW model was obtained in
general terms in \cite{ibsfclass}, and its specialization to particle
solutions was given explicitly in \cite{ibsfglo}.

A more detailed exploration of the general 2D classical string theory in any
curved spacetime (i.e. not only WZW models) was done in \cite{ibjs}. In 2D
the only non-trivial stringy solutions turn out to be necessarily folded
strings, and therefore they are the only path toward analyzing stringy
questions in a toy model. In addition to the interest in singular
gravitational behavior (such as black holes) there has also been a
long-standing interest in exploring consistent generalizations of
non-critical strings with the hope that they may be relevant for some branch
of physics. Folded strings fall into this category, especially in the area
of string-QCD relations. Therefore two aspects of string theory were
investigated: (i) strings in curved space-time and (ii) folded strings.

In papers \cite{ibjs}\cite{ddcl}\cite{cldd} the complete set of solutions of
two dimensional classical string theory were constructed for {\it any 2D
curved spacetime}. The classical action is given by $\smallint d^2\sigma
\,G_{\mu \nu }(x)\partial _{+}x^\mu \partial _{-}x^\nu .$ In 2D $B_{\mu \nu
}(x)$ can be eliminated since it produces a total derivative in the action,
and in the classical theory the dilaton is absent. The most general metric
can always be transformed into the conformal form $G_{\mu \nu }=\eta _{\mu
\nu }G(x).$ Then the most general 2D {\it classical} string equations of
motion and conformal (Virasoro) constraints take the form
\begin{equation}
\begin{array}{c}
\partial _{+}(G\,\,\partial _{-}u)+\partial _{-}(G\,\,\partial _{+}u)=\frac{%
\partial G}{\partial v}(\partial _{+}u\partial _{-}v+\partial _{+}v\partial
_{-}u) \\
\partial _{+}(G\,\,\partial _{-}v)+\partial _{-}(G\,\,\partial _{+}v)=\frac{%
\partial G}{\partial u}(\partial _{+}u\partial _{-}v+\partial _{+}v\partial
_{-}u) \\
\partial _{+}u\partial _{+}v=0=\partial _{-}u\partial _{-}v\,\,\,,
\end{array}
\label{stringeqs}
\end{equation}
where we have used the target space lightcone coordinates $\,\,u(\sigma
^{+},\sigma ^{-})=\frac 1{\sqrt{2}}(x^0+x^1),\,\,v(\sigma ^{+},\sigma
^{-})=\frac 1{\sqrt{2}}(x^0-x^1),$ and the world sheet lightcone coordinates
$\sigma ^{\pm }=(\tau \pm \sigma )/\sqrt{2},\,\,\,\partial _{\pm }=(\partial
_\tau \pm \partial _\sigma )/\sqrt{2}.$

In flat space-time the solutions are given in terms of arbitrary left-moving
and right-moving functions $x_L^\mu (\sigma ^{+}),x_R^\mu (\sigma ^{-})$
\[
x^\mu (\tau ,\sigma )=x_L^\mu (\sigma ^{+})+x_R^\mu (\sigma ^{-}).
\]
As shown by BBHP \cite{bbhp1}\cite{bbhp2}, the constraints are also
satisfied provided\footnote{%
Although the original BBHP solutions were for open strings, the same
solutions also apply to closed strings by simply taking independent
functions $f,g$ for left movers and right movers.}
\begin{equation}
\begin{array}{c}
u(\sigma ^{+},\sigma ^{-})=u_0+\frac{p^{+}}2\left[ (\sigma ^{+}+f(\sigma
^{+}))+(\sigma ^{-}-g(\sigma ^{-}))\right] \\
v(\sigma ^{+},\sigma ^{-})=v_0+\frac{p^{-}}2\left[ (\sigma ^{+}-f(\sigma
^{+}))+(\sigma ^{-}+g(\sigma ^{-}))\right]
\end{array}
\label{general}
\end{equation}
where $f(\sigma ^{+})$ and $g(\sigma ^{-})$ are any two {\it periodic
functions, } $f(\sigma ^{+})=f(\sigma ^{+}+\sqrt{2})$, $g(\sigma
^{-})=g(\sigma ^{-}+\sqrt{2}),$ with slopes $f^{\prime }(\sigma ^{+})=\pm 1$
and $g^{\prime }(\sigma ^{-})=\pm 1.$ The slope can change discontinuously
any number of times at arbitrary locations $\sigma _i^{+},\sigma _j^{-}$
within the basic intervals $-1/\sqrt{2}\leq \sigma ^{\pm }\leq 1/\sqrt{2}$
(and then repeated periodically), but the functions $f,g$ are continuous at
these points. The discontinuities in the slopes are allowed since the
equations of motion are first order in either $\partial _{+}$ or $\partial
_{-}$. The number of times the slope changes in the basic interval
corresponds to the number of folds for left movers and right movers
respectively. The simplest BBHP\ solution is the so called yo-yo solution
given by $f=|\sigma ^{+}|_{per}$ and $g=|\sigma ^{-}|_{per}$ which are the
periodically repeated absolute value. These solutions describe folded
strings, with the folds oscillating against each other, and moving at the
speed of light. Examples are plotted in Figures 1,2. In Fig.1 one sees the
yo-yo solution with equal periods for $|\sigma ^{+}|_{per}$, and $|\sigma
^{-}|_{per}$ . Fig. 2 is generated by taking the period of $|\sigma
^{-}|_{per}$ to be half of that of $|\sigma ^{+}|_{per}$ .

As discovered in {\cite{ibjs}\cite{ddcl}, }the complete set of classical
solutions in curved spacetime are classified by their behavior in the
asymptotically flat region of spacetime $G(u,v)\rightarrow 1,$ where they
tend to the folded string solutions of BBHP\ given in (\ref{general}) as
boundary conditions. The curved space solutions are given in the form of a
map from the world sheet to target spacetime, where (as a mathematical
convenience) the world sheet is divided into lattice-like patches
corresponding to different maps. The world-sheet lattice structure is
determined by the sign patterns of ($f^{\prime },g^{\prime })=(\pm ,\pm )$
inherent in the BBHP solutions, thus the lattice is dictated by the boundary
conditions in the asymptotically flat region of spacetime $G(u,v)\rightarrow
1$. The lattice is on the world-sheet, not in curved spacetime, it is only a
mathematical tool to keep track of patches, and the world sheet is not at
all discretized. In each patch of the lattice one set of signs holds, hence
there are 4 types of patches called $A,B,C,D.$ For each such patch there is
a solution of the equations of motion that is valid within the patch. The
forms of the solutions in patches labelled by an integer $k$ are (see eq.(%
\ref{soll}) for an example of a pattern of patches)
\begin{equation}
\begin{array}{ll}
A:\qquad u=U_k(\sigma ^{+}), & \quad v=V_k(\sigma ^{-}) \\
B:\qquad u=U_k(\sigma ^{-}), & \quad v=V_k(\sigma ^{+}) \\
C:\qquad u=u_k, & \quad v=W[\alpha _k(\sigma ^{+})+\beta _k(\sigma
^{-}),\,\,u_k] \\
D:\qquad u=\bar W[\alpha _k(\sigma ^{-})+\beta _k(\sigma ^{+}),\,\,v_k], &
\quad v=v_k\,\,\,,
\end{array}
\label{fourr}
\end{equation}
where the constants $u_k,v_k$ and the functions $U_k(\sigma ^{\pm
}),V_k(\sigma ^{\pm })$, $\alpha _k(\sigma ^{\pm })$, $\beta _k(\sigma ^{\pm
})$ are given by a recursion relation whose form depends on the metric $G.$
It is easy to verify that, independently of the recursion relation, the
forms listed in (\ref{fourr}) solve the differential equations for any $%
U_k(\sigma ^{\pm }),V_k(\sigma ^{\pm })$, $\alpha _k(\sigma ^{\pm })$, $%
\beta _k(\sigma ^{\pm })$ provided the functions $W,\bar W$ are defined by
inverting the following functions
\[
\int^Wdv^{\prime }G(u_k,v^{\prime })=\alpha +\bar \beta ,\quad \,\int^{\bar
W}du^{\prime }G(u^{\prime },v_k)=\bar \alpha +\beta .
\]
By construction, in flat spacetime the recursion reproduces the BBHP
solutions given above.

The recursion relation, which is analogous to a ``transfer matrix'',
connects the maps in different patches into a single continuous map. It is
derived by demanding continuity accross the boundaries of each patch (see
below for an example). Thus, the functions in the various patches get
related to each other. This ``transfer matrix'' encodes the properties of
the world sheet lattice on the one hand and the geometry of spacetime on the
other hand. {Thus, }lattices on the world-sheet plus geometry in space-time
lead to ``transfer matrices''. Recall that the lattice is dictated by the
nature of the solution (\ref{general}) in the asymptotically flat region of
target spacetime. This seems to be a rich area of mathematical physics to
explore in more detail in the future.

As an example we consider the simplest yo-yo solution as a boundary
condition. This defines the sign patterns according to the slopes of the
periodic functions $|\sigma ^{+}|_{per}$ and $|\sigma ^{-}|_{per},$ and the
following lattice emerges from the periodic behavior of these functions .
The world{\ \ sheet is labelled by $\sigma $ horizontally and by $\tau $
vertically. Periodicity in }$\sigma $ is imposed, hence the world sheet is a
cylinder. {It is sliced by equally spaced $45^o$ lines that form a
light-cone lattice in $\sigma ^{\pm }$. The crosses in the diagram represent
the corners of the cells on the world sheet.} {\tiny
\begin{equation}
\begin{array}{ccccc}
\begin{array}{c}
\sigma =0 \\
\vdots
\end{array}
&
\begin{array}{c}
\sigma =1 \\
\vdots
\end{array}
&
\begin{array}{c}
\sigma =2 \\
\vdots
\end{array}
&
\begin{array}{c}
\sigma =3 \\
\vdots
\end{array}
&
\begin{array}{c}
\sigma =4\equiv 0 \\
\vdots
\end{array}
\\
\times &
\begin{array}{c}
\begin{array}{c}
U_{k+2}(\sigma ^{+}) \\
V_{k+2}(\sigma ^{-})
\end{array}
\end{array}
& \times &
\begin{array}{c}
U_{k+2}(\sigma ^{-}) \\
V_{k+2}(\sigma ^{+})
\end{array}
& \times \\
&  &  &  &  \\
\cdots
\begin{array}{c}
u_{k+1} \\
W_{k+1}(\sigma ^{+},\sigma ^{-})
\end{array}
& \times &
\begin{array}{c}
\bar W_{k+1}(\sigma ^{+},\sigma ^{-}) \\
v_{k+1}
\end{array}
& \times &
\begin{array}{c}
u_{k+1} \\
W_{k+1}(\sigma ^{+},\sigma ^{-})
\end{array}
\cdots \\
&  &  &  &  \\
\times &
\begin{array}{c}
U_{k+1}(\sigma ^{-}) \\
V_{k+1}(\sigma ^{+})
\end{array}
& \times &
\begin{array}{c}
U_{k+1}(\sigma ^{+}) \\
V_{k+1}(\sigma ^{-})
\end{array}
& \times \\
&  &  &  &  \\
\cdots
\begin{array}{c}
\bar W_k(\sigma ^{+},\sigma ^{-}) \\
v_k
\end{array}
& \times &
\begin{array}{c}
u_k \\
W_k(\sigma ^{+},\sigma ^{-})
\end{array}
& \times &
\begin{array}{c}
\bar W_k(\sigma ^{+},\sigma ^{-}) \\
v_k
\end{array}
\cdots \\
&  &  &  &  \\
\times &
\begin{array}{c}
U_k(\sigma ^{+}) \\
V_k(\sigma ^{-})
\end{array}
& \times &
\begin{array}{c}
U_k(\sigma ^{-}) \\
V_k(\sigma ^{+})
\end{array}
& \times \\
&  &  &  &  \\
\cdots
\begin{array}{c}
u_{k-1} \\
W_{k-1}(\sigma ^{+},\sigma ^{-})
\end{array}
& \times &
\begin{array}{c}
\bar W_{k-1}(\sigma ^{+},\sigma ^{-}) \\
v_{k-1}
\end{array}
& \times &
\begin{array}{c}
u_{k-1} \\
W_{k-1}(\sigma ^{+},\sigma ^{-})
\end{array}
\cdots \\
\quad \vdots & \vdots & \quad \quad \vdots \quad \quad & \vdots & \quad
\vdots \quad
\end{array}
\label{soll}
\end{equation}
}

The transfer matrix for this ``yo-yo lattice'' was derived in {{\cite{ibjs}%
\cite{cldd}\cite{ddcl} for any metric }}$G${.} Here we give only the results
for the $SL(2,R)/R$ {black hole space-time }$ds^2=du\,dv(1-uv)^{-1}$ {{and
for the cosmological deSitter space-time {${{{{ds^2=dt^2-R^2(t)\frac{dr^2}{%
1-kr^2}=\frac{4\,}{H^2}(u+v)^{-2}\,}}}}du\,dv{,}$}}} for {{{${\,}$ ${{{\,}%
|R(t)|=e^{Ht}}},$ }}}$k{=0}${. }

For the black hole metric the ``transfer matrix'' is{{\
\begin{equation}
\begin{array}{c}
\bar W_k=\frac 1{v_k}\left[ 1-\frac{\left( 1-U_k(\sigma ^{+})v_k\right) \
\left( 1-U_k(\sigma ^{-})v_k\right) }{1-u_{k-1}v_k}\right] \\
U_{k+1}(z)=\frac{1-u_kv_k}{1-u_{k-1}v_k}\left[ U_k(z)+\frac{u_k-u_{k-1}}{%
1-u_kv_k}\right] \\
u_{k+1}=\frac{2u_k-u_{k-1}-u_k^2v_k}{1-u_{k-1}v_k},
\end{array}
\label{recsl}
\end{equation}
and similarly $W_k,V_k,v_k$ are obtained from the above by interchanging $%
U\leftrightarrow V$ and $u\leftrightarrow v.$ The constants }}$u_k,v_k$ are
the values of the functions $U_k(z),V_k(z)$ at the boundaries of the cell
labelled by $k:$%
\[
u_{k-1}=U_k(-1/\sqrt{2}),\quad u_k=U_k(1/\sqrt{2}),\quad etc.
\]
These constants describe the motion of folds that move at the speed of
light. {Note that for $u,v\rightarrow 0$ or $\infty $ the metric approaches
the flat metric. }

{Remarkably, there is an invariant of this ``transfer matrix'' that
corresponds to the ``lattice area'' swept by the string [for comparison,
recall the form of the action density whose meaning is area }$%
dA=(1-uv)^{-1}(\partial _{+}u\partial _{-}v+\partial _{+}v\partial _{-}u)]$

{{\
\begin{equation}
dA_k=\frac{(u_k-u_{k-1})\,(v_k-v_{k-1})}{1-\frac
14(u_k+u_{k-1})\,(v_k+v_{k-1})}.  \label{minimsl}
\end{equation}
It can be easily verified that }}$dA_{k+1}=dA_k$, implying that this
quantity remains a constant even in the vicinity of singularities $uv\approx
1$. This observation leads to new interesting phenomena as discussed below.

The ``transfer matrix'' for the deSitter spacetime is{\ {\
\begin{equation}
\begin{array}{c}
\bar W_k(\sigma ^{+},\sigma ^{-})=\left[ \frac 1{U_k(\sigma ^{+})+v_k}+\frac
1{U_k(\sigma ^{-})+v_k}-\frac 1{u_{k-1}+v_k}\right] ^{-1}-v_k \\
U_{k+1}(z)=\left[ \frac 1{U_k(z)+v_k}+\frac 1{u_k+v_k}-\frac
1{u_{k-1}+v_k}\right] ^{-1}-v_k \\
u_{k+1}=\left[ \frac 2{u_k+v_k}-\frac 1{u_k+v_{k-1}}\right] ^{-1}-v_k
\end{array}
\label{transdesit}
\end{equation}
Similar formulas hold for $W_k,V_k,v_k$ respectively. In this case too there
is an invariant area }}
\[
dA_k=\frac 4{H^2}\frac{\left( u_k-u_{k-1}\right) \left( v_k-v_{k-1}\right) }{%
\left( u_k+v_{k-1}\right) \left( u_{k-1}+v_k\right) }
\]
{{The 2D deSitter space can be embedded in 3D as the surface of a
hyperboloid described by
\begin{equation}
\begin{array}{c}
x_0^2-x_1^2-x_2^2=-H^{-2} \\
x_0=\frac{uv-H^{-2}}{u+v},\quad x_1=\frac{uv+H^{-2}}{u+v},\quad x_2=\frac 1H%
\frac{u-v}{u+v}
\end{array}
\label{D23D}
\end{equation}
Then the deSitter metric}} {{takes the flat form
\begin{equation}
ds^2=dx_0^2-dx_1^2-dx_2^2.  \label{flatdesit}
\end{equation}
The motion is more easily visualized in this parametrization.}}

The recursion relations are solved in terms of two functions $U_0(z),V_0(z)$
that are associated with the initial cell $k=0.$ {{The remaining conformal
invariance may be used to fix the form of the functions {$U_0(z),\,V_0(z)$ }%
in the initial cell (although this is not necessary). For the yo-yo solution
the initial functions $U_0(z),\,V_0(z)$ need not contain more than $4$
constants that are related to the initial positions and velocities of the
two folds. However, there is a physical requirement: the time coordinate }}$%
x^0(\tau ,\sigma )$ constructed from {{\ $U_0(\sigma ^{\pm }),\,V_0(\sigma
^{\pm })$ }}must be an increasing function of the proper time $\tau $ for
any value of $\sigma ,$ so that physically every point on the string moves
forward in time (no bits of anti-strings). {{Therefore, the simplest
physical gauge fixed form is
\begin{equation}
\begin{array}{c}
U_0(z)=\frac 12(u_0+u_{-1})+\frac 1{\sqrt{2}}(u_0-u_{-1})\,z_{per} \\
V_0(z)=\frac 12(v_0+v_{-1})+\frac 1{\sqrt{2}}(v_0-v_{-1})\,z_{per},
\end{array}
\label{initial}
\end{equation}
where $z_{per}$ is the linear function $z_{per}=z$ in the interval $-1/\sqrt{%
2}\leq z\leq 1/\sqrt{2}$, and then repeated periodically. This form
reproduces the BBHP yo-yo solution in flat spacetime from the recursion
relations, provided one uses }}$G(u,v)=1.$ {However, any other function with
the same 4 boundary constants and general increasing character will produce
the same, gauge independent, physical motion for the folds in flat or curved
spacetime, since their motion is given by gauge independent equations
involving only the gauge independent constants }$u_k,v_k${. Evidently, the
motion of the intermediate points of the string is gauge dependent, as
expected.}

{\ {The constants $(u_k,v_k)$ are sufficient to describe the physical motion
of the folds (or end points), as well as the whole string. }T{he
trajectories of the folds are plotted in Figs.4,5,6 {\ by feeding the
recursion relations to a computer}. }}As in Fig.3, {the string performs
oscillations that are similar to those of flat spacetime around a center of
mass that follows on the average the geodesic of a massive point particle.
This is expected intuitively. A detailed discussion of the black hole case
was given in {\cite{ibjs}.} The main surprize is the tunelling of the string
into the forbidden region in Fig.5 (the bare singularity region of the black
hole), where particles cannot go. This behavior cannot be avoided since it
follows from minimal area {\it conservation laws} that were given above {%
\cite{ibjs}{\cite{ddcl}}\cite{cldd}}. It could be compared to diffraction,
or light illuminating the wall around a corner, that can happen with waves,
but not with particles. In addition, the {\it massive} point particle
geodesic (as well as the string geodesic) does not stop at the black hole,
rather it reaches the black hole in a finite amount of proper time, and then
it continues into a second sheet of spacetime that is glued to the first
sheet at the black hole singularity. The observers on the second sheet see
it as if the massive particle or the string is coming out of a white hole.
The motion may continue from white hole to black hole singularities, each
time moving into a new sheet, interpreted as a new world, like in the
Reissner-Nordstrom spacetime. For more details see {\cite{ibjs}{\cite{ddcl}}%
\cite{cldd}.}}

{\ }

\section{Quantum Folded String}

{\ Given the fact that the string in 2D is quite non-trivial classically, we
expect that there is a consistent quantization procedure that includes the
non-trivial folded states. Therefore we should try to make a case for folded
strings in the quantum theory.}

As pointed out many times in our past work, folded 2D-string states are
present in the $d=2$ and $c\leq 25$ sector of the quantum theory in flat as
well as curved spacetime. In simple string models, when it has been possible
to compute the spectrum, their norm is positive and is proportional to $%
(c-26)$. Only if $d=2$ and $c=26$ simultaneously (e.g. $d=2$ flat space-time
with linear dilaton such that $c=26$) the folded string states become zero
norm states and then the special discrete momentum states survive as the
only stringy states. A simple model in which these properties may be easily
seen is the {\it covariant} quantization of the 2D string theory, in which
the physical states are identified as the subset that satisfies the Virasoro
constraints, i.e. $L_0-\frac{d-2}{24}=L_{n_{\geq 1}}=0$ applied on states.
For example, it has been known for a long time that the $d\leq 25$ sector of
the flat string theory has non-trivial positive norm states (including for $%
d=2)$ that satisfy the Virasoro constraints and that there are no ghosts
\cite{noghost}. {A similar covariant quantization can be carried out for the
2D black hole string by using the Kac-Moody current algebra formulation, and
relaxing the $c=26$ condition (i.e. $k<9/4$) to include the folded strings. }

Why $c=26?$ There are several approaches to the quantization of strings that
converge on the requirement of $c=26.$ These include the light-cone gauge,
the Polyakov path integral and the BRST quantization. However, they each
involve certain steps that seem to inadverdently exclude the $c<26$ string.
We can point out that

\begin{description}
\item  (i) The usual light-cone approach throws away the folded states from
the beginning by assuming a uniform momentum density $P^{+}(\tau ,\sigma
)=p^{+}$, a statement that is not true for the BBHP solutions even in flat
spacetime.

\item  (ii) The Polyakov approach assumes a certain measure for the path
integral, thus locking into a {\it definition} of a quantum theory. A
different measure that takes into account folds can be considered as in {{%
\cite{yankielowicz} mentioned below.}}

\item  {(iii) the BRST approach requires }${Q}_{BRST}^2=0$ as an operator.
This is a stronger requirement than the Virasoro constraints satisfied only
in the physical subspace $<phys|L_n-\alpha _0\delta _{n0}|phys>=0$. An
analogous statement would be $<phys|Q_{BRST}|phys>=0,$ which does not lead
to $c=26.$ Actually, the fact that there exists a consistent covariant
quantization of the flat free string in $d<26$ is already proof that the $%
Q_{BRST}^2=0$ approach is too strong.
\end{description}

\noindent Therefore, it appears that a more general quantization of string
theory for $c<26$, that would permit folded string states, seems possible. {%
What would also be interesting is to find the correct formulation for
interacting folded strings. The path integral approach }discussed{{\ in \cite
{yankielowicz} seems to be promising, and it may be possible to make faster
progress by reformulating it in the conformal gauge and relating it to our
classical solutions \ Note that the definition of fold in ref.\cite
{yankielowicz} does not take into account that the map from the world sheet
to spacetime may be many to one (i.e. a region maped to a segment, as is the
case for our solutions)}}. {{This feature may be important in the
formulation of folds and their interactions in the path integral approach.
In particular, the description of folds in the conformal gauge, as in our
papers, may eventually prove to be a more convenient mathematical
formulation than the one used in \cite{yankielowicz}.}}

\section{Higher dimensions}

{\ {\ Folded strings exist in higher dimensions as well. One can display the
general solution in flat space-time in the temporal gauge
\begin{equation}
\begin{array}{c}
x^0=p^0\tau ,\quad \vec x(\tau ,\sigma )=\vec x_L(\sigma ^{+})+\vec
x_R(\sigma ^{-}),\quad (\partial _{+}\vec x_L)^2=p_0^2=(\partial _{-}\vec
x_R)^2 \\
\\
\partial _{+}\vec x_L=p^0\left( \frac{2{\bf f}}{1+{\bf f}^2},\frac{1-{\bf f}%
^2}{1+{\bf f}^2}\varepsilon _L\right) ,\quad \partial _{-}\vec x_R=p^0\left(
\frac{2{\bf g}}{1+{\bf g}^2},\frac{1-{\bf g}^2}{1+{\bf g}^2}\varepsilon
_R\right)
\end{array}
\label{dfolded}
\end{equation}
where ${\bf f}(\sigma ^{+}){\bf ,\,\,g(\sigma ^{-})}$ are arbitrary periodic
vectors in $d-2$ dimensions, {\it which could be discontinuous}, and $%
\varepsilon _L(\sigma ^{+}),\,\varepsilon _R(\sigma ^{-})$ take the values $%
\pm 1$ in patches of the corresponding variables such that the sign patterns
repeat periodically (as in the 2D\ string). When ${\bf f,g}$ are both zero
the solution reduces to the 2 dimensional BBHP case. In general, the
presence of discontinuous $\varepsilon _{L,}\varepsilon _R,$ and the
discontinuities in ${\bf f}(\sigma ^{+}){\bf ,\,\,g(\sigma ^{-})}$ gives a
larger set of solutions, which include strings that are partially or fully
folded. Discontinuities are allowed since the differential equations are
first order in the derivatives }$\partial _{+}$ and $\partial _{-}.${\ }Such
solutions are usually missed in the lightcone gauge even in the flat
classical theory (therefore, the lightcone ``gauge'' is not really a gauge).
}

{\ The curved space-time analogs of such solutions in higher dimensions are
presently under investigation. }

{\ }

\section{Comments and Conclusions}

{\ {\ We have solved generally the classical 2D string theory in any curved
space-time. All stringy solutions correspond to folded strings. All
solutions tend to the BBHP\ solutions (as boundary conditions) in the
asymptotically flat region of the curved space-time. Therefore, the BBHP
solutions of eq.(\ref{general}) serve to classify all the solutions for any
curved space-time. In fact, the sign patterns of the BBHP solutions provide
the method for dividing the world-sheet into patches, thus defining the
lattices associated with the $A,B,C,D$ solutions. The matching of boundaries
for these functions gives the general solution in curved space-time in the
form of a ``transfer matrix''. Thus, }}lattices on the world-sheet plus
geometry in space-time lead to transfer matrices. This seems to be a rich
area to explore in more detail.

{\ {\ The general physical motion of the string is: oscillations around a
center of masss that follows on the average a geodesic of a massive
particle, consistent with intuition. The oscillations are deformed by
curvature as compared to the BBHP solutions in flat spacetime, but they
maintain the same general character as long as the curvature is smooth.
However, new stringy behavior becomes evident in the vicinity of
singularities where new phenomena, such as tunneling (similar to
diffraction), take place. } There is also the continuation of the motion
into new worlds, in a finite amount of proper time, that the string as well
as the massive particle geodesics do (but not the massless particle! - see
\cite{ddcl}\cite{cldd}). Because of the tunelling and the new worlds, the
global space of the }$SL(2,R)/R$ black hole is not just the usual black hole
space, $uv<1.$ Rather, it must include also the $uv>1$ ``bare singularity''
region even for the classical description of strings (actually this region
is not really singular, as argued in \cite{ddcl}\cite{cldd}). We conjecture
that the inclusion of the bare singularity region is a more general
requirement than the $SL(2,R)/R$ case for the correct description of string
motion. Of course, by duality, the quantum theory must include all the
regions.

Folded string are also of interest in a string-QCD relation. Gluons are
expected to behave just like the folds, since only at the location of a
gluon the color flux tube can fold. Some recent discussion if this point can
be found in \cite{ibberkeley} and \cite{ddcl}\cite{cldd}.

{We suspect that the inclusion of the quantum states corresponding to folded
strings may lead to a consistent quantum theory in less than 26 dimensions.
As already emphasized earlier in the paper, the free string is perfectly
consistent as a quantum theory for $c<26$, including the folded states. The
interacting quantum string with folds remains as an open possibility.}



\newcommand\putfig[3]{
   \vbox{
   \let\picnaturalsize=N
   \def\picsize{#3}
   \def\picfilename{#1}
   \ifx\nopictures Y\else{\ifx\epsfloaded Y\else\input epsf \fi
   \let\epsfloaded=Y
   \centerline{\ifx\picnaturalsize N\epsfxsize \picsize\fi
   \epsfbox{\picfilename}}}\fi
   \vspace{1.0cm}
   {\it #2}
   \vspace{1.5cm}
   }
}

\begin{figure}
\putfig{ 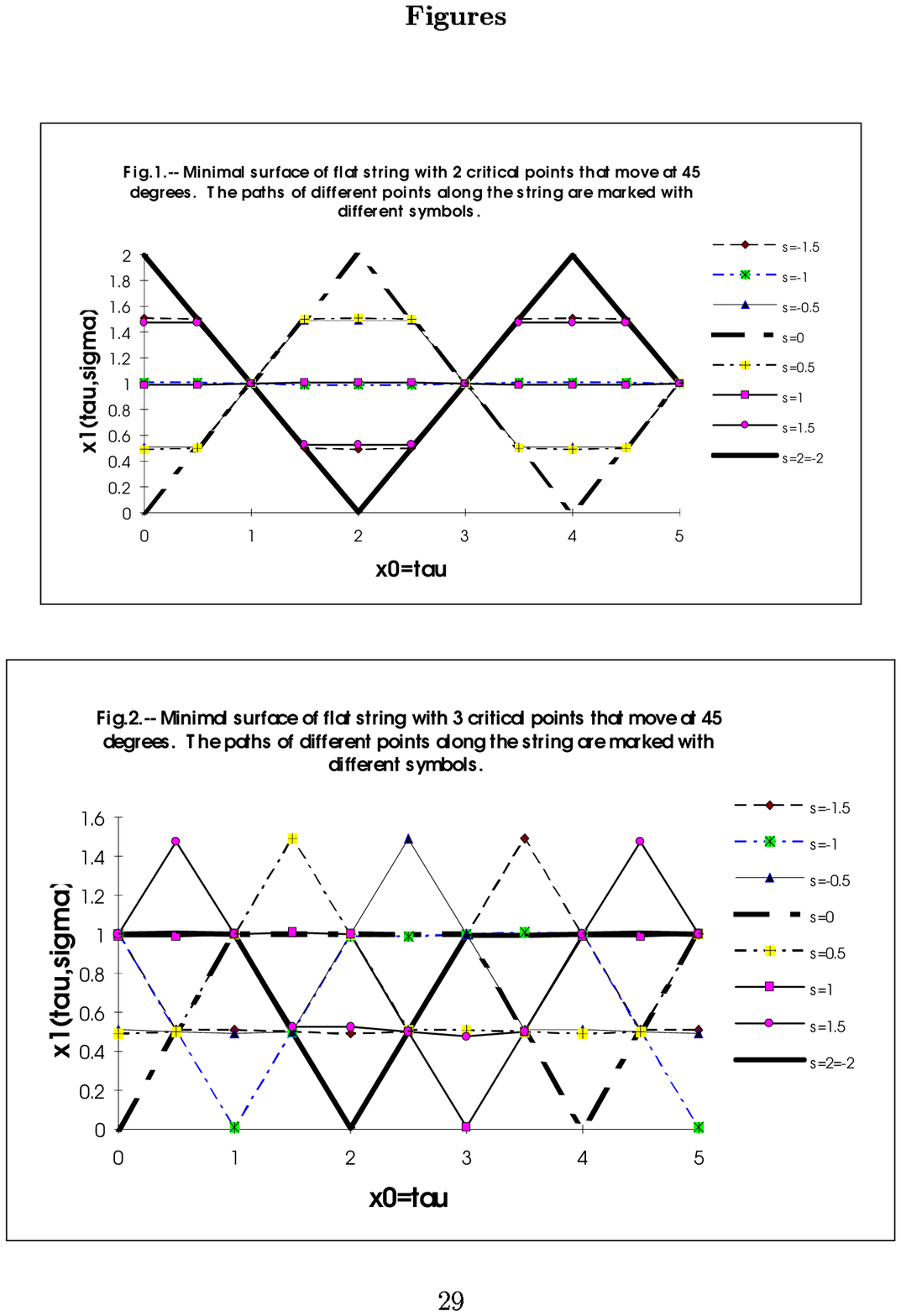}{150mm}{200mm}
\end{figure}

\begin{figure}
\putfig{ 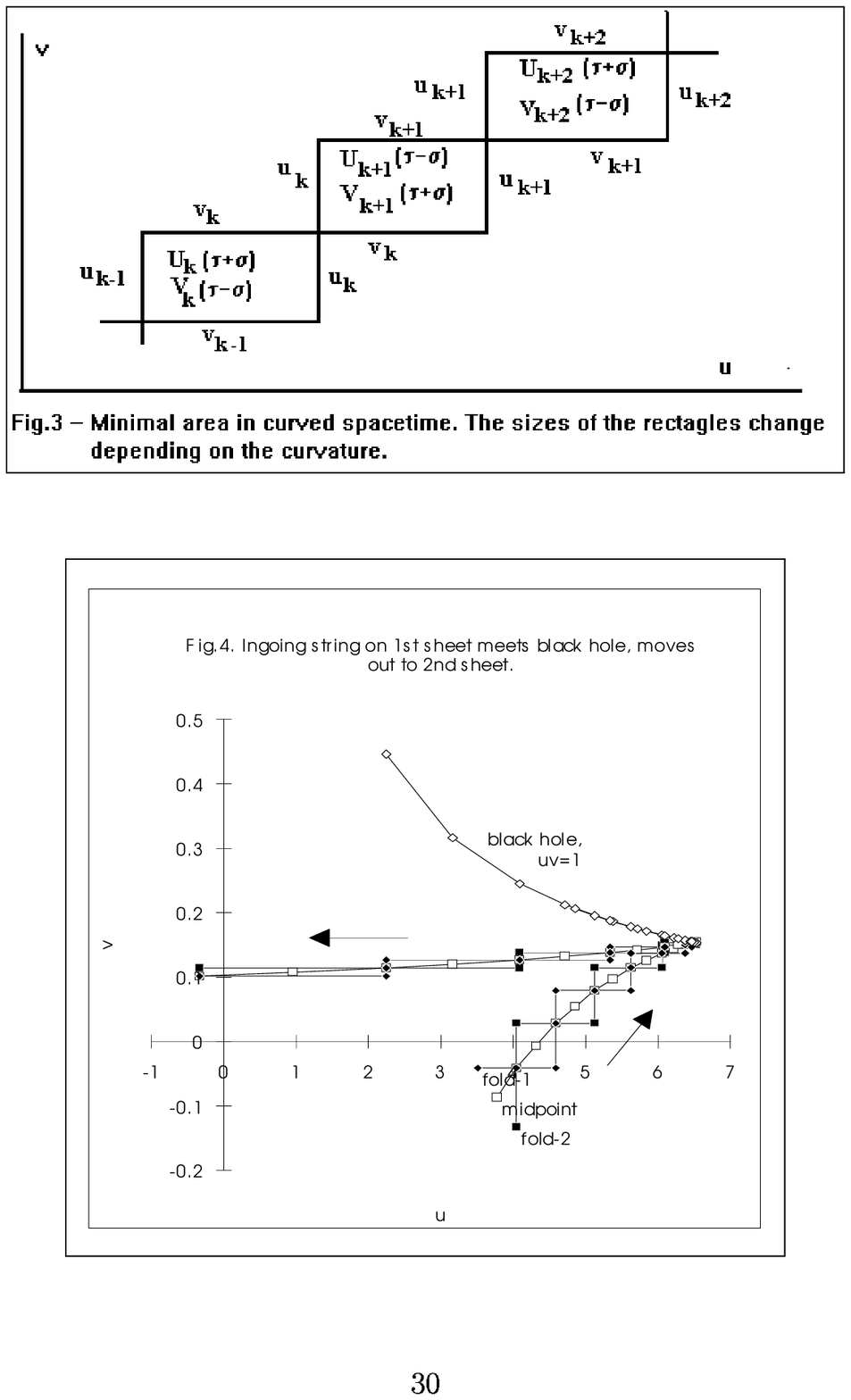}{150mm}{200mm}
\end{figure}

\begin{figure}
\putfig{ 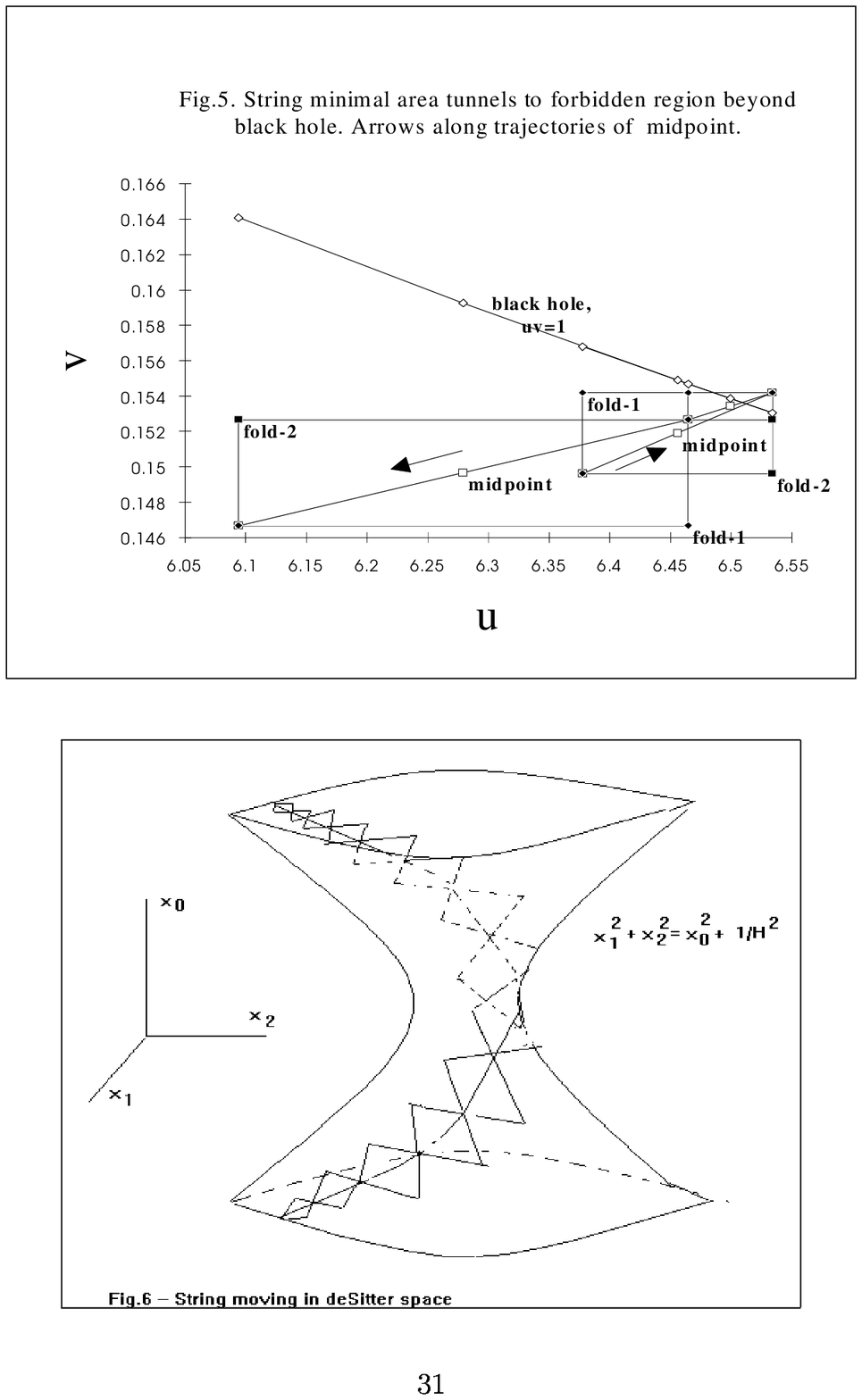}{150mm}{200mm}
\end{figure}



\end{document}
